\documentclass{article}
\usepackage{spconf}

\usepackage{cite,hyperref,balance}
\usepackage{amsmath,amssymb,amsfonts}
\usepackage{algorithmic}
\usepackage{graphicx,booktabs,multirow}
\usepackage{textcomp}
\usepackage{xcolor}

\begin{document}
\ninept

\title{
A Hierarchical Regression Chain Framework\\
for Affective Vocal Burst Recognition
}

\name{
Jinchao Li$^1$, Xixin Wu$^{1*}$, Kaitao Song$^2$, Dongsheng Li$^2$, Xunying Liu$^1$, Helen Meng$^1$
\thanks{$^1$The work was done during the author’s internship with Microsoft. $^*$Corresponding author.}
}
\address{
$^1$The Chinese University of Hong Kong, Hong Kong SAR, China\\
$^2$Microsoft Research Asia, Shanghai, China\\
{\tt \small $^1$\{jcli,wuxx,xyliu,hmmeng\}@se.cuhk.edu.hk, $^2$\{kaitaosong,Dongsheng.Li\}@microsoft.com}
}

\maketitle

\begin{abstract}
%Vocal bursts, non-linguistic vocalizations, play important roles in robust and general speech emotion recognition. We present our approach for modeling affective vocal bursts in the ACII Affective Vocal Burst Workshop \& Challenge 2022 (A-VB).
%The proposed methods use self-supervised acoustic model to extract features, and a hierarchical bi-chain regression framework to model the labels' dependency.
%Experimental results show the effectiveness of proposed components, and give a superior performance compared to several baseline methods, e.g., mean concordance correlation coefficients of 0.6854, 0.7237, 0.6017 for the TWO, HIGH and CULTURE tasks.
As a common way of emotion signaling via non-linguistic vocalizations, vocal burst (VB) plays an important role in daily social interaction. Understanding and modeling human vocal bursts are indispensable for developing robust and general artificial intelligence. Exploring computational approaches for understanding vocal bursts is attracting increasing research attention. In this work, we propose a hierarchical framework, based on chain regression models, for affective recognition from VBs, that explicitly considers multiple relationships: (i) between emotional states and diverse cultures; (ii) between low-dimensional (arousal \& valence) and high-dimensional (10 emotion classes) emotion spaces; and (iii) between various emotion classes within the high-dimensional space. To address the challenge of data sparsity, we also use self-supervised learning (SSL) representations with layer-wise and temporal aggregation modules. The proposed systems participated in the ACII Affective Vocal Burst (A-VB) Challenge 2022 and ranked first in the ``TWO'' and ``CULTURE'' tasks. Experimental results based on the ACII Challenge 2022 dataset demonstrate the superior performance of the proposed system and the effectiveness of considering multiple relationships using hierarchical regression chain models.

\end{abstract}

\section{Introduction}
\label{sec:into}
% Background (VB, competition) -> Research Gap -> Motivation -> Our work description

Recognition of emotions conveyed by non-linguistic vocalizations, e.g., affective bursts, has attracted increasing research attention, as vocalizations can reliably express certain emotions and the meanings of vocal bursts are generally preserved across diverse cultures \cite{cowen2019mapping}.
Similar to facial expressions, the affective vocalization information lays the foundation of more robustly and holistically understanding emotional reactions.
Despite the fact that affect vocalizations and speech-embedded prosody both utilize the same expressive (vocal) apparatus, it is also found that the accuracy of emotion decoding for non-linguistic affect vocalizations is higher than the accuracy for speech-embedded vocal prosody~\cite{hawk2009worth}. 
% \cite{scherer1986vocal,hawk2009worth,simon2009voice}.
% necessary of research on AVB
Much research has been conducted in speech emotion recognition (SER) with verbal speech recently, such as feature exploration~\cite{liang2021multibench,li2022context} and multilingual generalization~\cite{singh2022systematic}.
% \textcolor{red}{related works on tradition vocal features, and multilingual ER}
Although some research investigated the combination of verbal and non-verbal speech~\cite{huang2019speech,hsu2021speech}, emotion recognition with nonverbal vocalizations only have still been received less attention.

Due to the scarcity of vocal burst data and lack of understanding about mechanisms of emotion signaling via vocal bursts, developing computational models for such emotion signaling remains a challenging task.   
Therefore, the recent ICML Expressive Vocalisations Workshop \& Competition 2022 (ExVo) and the ACII Affective Vocal Bursts Workshop \& Challenge 2022 (A-VB) introduce the large-scale Hume Vocal Bursts Competition Dataset (HUME-VB) for exploring various computational approaches \cite{baird2022icml,baird2022acii}. The corpus contains about 37 hours of self-recorded data by speakers in 4 countries spanning 3 native languages, which can support investigation of affective vocal bursts from diverse perspectives. Multi-task approaches have been demonstrated to be effective in previous works, e.g., by integrating various losses \cite{jing2022redundancy}, jointly modeling auxiliary prediction tasks of culture and age \cite{song2022dynamic}. However, it is desirable to explicitly model the relationship between emotion classes in vocalization signaling and the relationship between the different related tasks.
% Cowen2022HumeVB,BairdA-VB2022,
To this end, we propose a hierarchical framework based on chain regression models, which generate predictions for one task that is conditioned on the prediction from the other related tasks.

% related works: features, methods (multitask)
With recent advancements in self-supervised learning (SSL), the adopted speech representations for emotion recognition are shifting from hand-crafted features, e.g., acoustic pitch and energy, to high-level embeddings extracted by pretrained models, such as Wav2vec 2.0 \cite{baevski2020wav2vec}. The large, Transformer-based SSL models trained on large-scale data can learn representations for various downstream tasks, including automatic speech recognition (ASR) \cite{baevski2020wav2vec} and SER \cite{sharma2022multi}. 
As affective vocal burst (AVB) data is generally lacking, it is important to borrow data from other speech domains to improve the AVB modeling. 
%better and more generalizable representations that are critical for extracting the affective cues. 
Purohit \textit{et al.}~\cite{purohit2022comparing} compared supervised and self-supervised embeddings for the affective vocal burst recognition (AVBR), and showed that SSL-based representations typically yield better performance than supervised embeddings learned by pretrained task-dependent neural networks.
To further leverage these high-level features, various network architectures have been explored in the latest SER research, such as layer-wise aggregation~\cite{pepino2021emotion}, temporal attention~\cite{liu2020temporal}, dynamic convolution~\cite{wen2021crossmodal}, and multi-labeling~\cite{chochlakis2022leveraging}.
Following \cite{pepino2021emotion}, we leverage representations from different layers of pretrained models with trainable weights.
% xin2022exploring

In this work, we investigate AVBR based on a hierarchical framework using chain regression models and pretrained representations. The relationships between emotional states and diverse cultures, between low-dimension and high-dimension emotion spaces, and between various emotion classes within the high-dimension space, are explicitly modeled. 
% Multiple classifiers and regressors are \textcolor{red}{XXX}. 
Our system participated in the ACII A-VB challenge and ranked first in the task of the ``TWO'' and ``CULTURE'' tasks, and second in the ``HIGH'' task.
The effectiveness of the regression models and the weighted aggregation of pretrained representations is also demonstrated by further experiments we conduct on the challenge dataset.

\section{Methodology}
\label{sec:method}
\begin{figure}[htb]
    \centering
        \includegraphics[width=0.9\columnwidth]{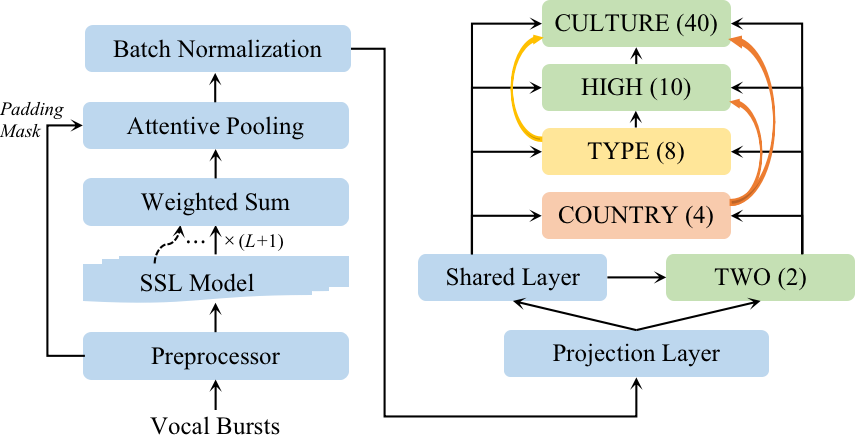}
    \vspace{-1em}
    \caption{Overview of hierarchical multitask learning framework for the A-VB 2022 competition. ``TWO'', ``TYPE'', ``HIGH'' and ``CULTURE'' denote the classifiers or regression models for corresponding tasks, and ``COUNTRY'' is the classifier of countries. ``$(*)$'' means the output size of the models.}
    \label{fig:overview}
    \vspace{-1em}
\end{figure}

% Hierarchical MTL architecture
The proposed hierarchical multitask learning framework is illustrated in Fig.~\ref{fig:overview}, mainly consisting of a high-level feature extractor (see left side Fig.~\ref{fig:overview}), and a structured output layer with a bi-directional regression chain (see right side Fig.~\ref{fig:overview}).
In the following, we will describe our proposed framework from the bottom levels of feature extraction, to the representation aggregation across different pre-trained model layers, and to the top structured output layer. 
%It is mainly composed In the following, we will describe each component separately.

\subsection{Preprocessing}
We preprocess the vocal burst data with peak normalization in the time domain for internal consistency of the data.
In addition, we use data augmentation to enrich the data and improve the robustness by slightly changing the acoustic characteristics with minor distortions. Specifically, we applied pitch-shifting and speed-perturbation for each input waveform during the training stage (corpus size not changed)~\cite{cariani1996neural,colosi1998efficient}.
The shifted ranges of pitch and speed are [-100, 100] semitones and [-0.05, +0.05] rates, respectively.

\subsection{High-level Feature Extractor}
% wav2vec2, weighted sum, attentive pooling, batch norm, proj, shared -> cnt, two, type, high, culture
The recent success of large pre-trained models motivates this work to adopt hidden embeddings from SSL models~\cite{adoma2020comparative,wang2021fine}.
We use the Wav2vec 2.0-Large XLSR (``w2v2-lg-xlsr'')~\cite{baevski2020wav2vec} to extract cross-lingual contextualised speech representations~\cite{conneau2020unsupervised}. The Wav2vec 2.0 Large XLSR is trained on the CommonVoice corpus~\cite{ardila2019common} by solving a contrastive task over masked latent speech representations and jointly learning a quantization of the latent representations shared across various languages.

The ``w2v2-lg-xlsr'' model contains one convolutional feature encoding layer and 24 stacked Transformer layers.
The convolutional layer contains temporal convolutions with kernel widths (10,3,3,3,3,2,2) and strides (5,2,2,2,2,2,2), which yield a receptive field of about 320 samples. Through this convolutional layer, we can obtain a feature map with a shape of $49\times1024$ (dimensions of time and the embedding, respectively) for each one-second segment with a 16kHz sampling rate from input vocal burst signals. To avoid information loss caused by only using the last Transformer layer of the Wav2vec model, we leverage both the Transformer layers and the convolutional layer. We use learnable weights to sum up all the hidden states of the 24 stacked Transformer layers and the output feature map from the convolutional layer.

An attentive time pooling layer~\cite{santos2016attentive} follows the weighted summed features and is used to compress the feature sequence with variable lengths into a fixed-length vector. The attention mechanism also enables flexible focus on important frames for target prediction, by allocating more weights to the corresponding frames in the summation. Then, we project the features into a lower-dimensional vector to reduce redundancy, while retaining the intra-class variability. A batch normalization layer \cite{ioffe2015batch} is applied to standardize the high-level features before the features are fed to the subsequent classifiers and regressors.

\begin{figure}[htb]
\centering
\begin{minipage}[b]{.9\linewidth}
    \centerline{\includegraphics[width=\linewidth]{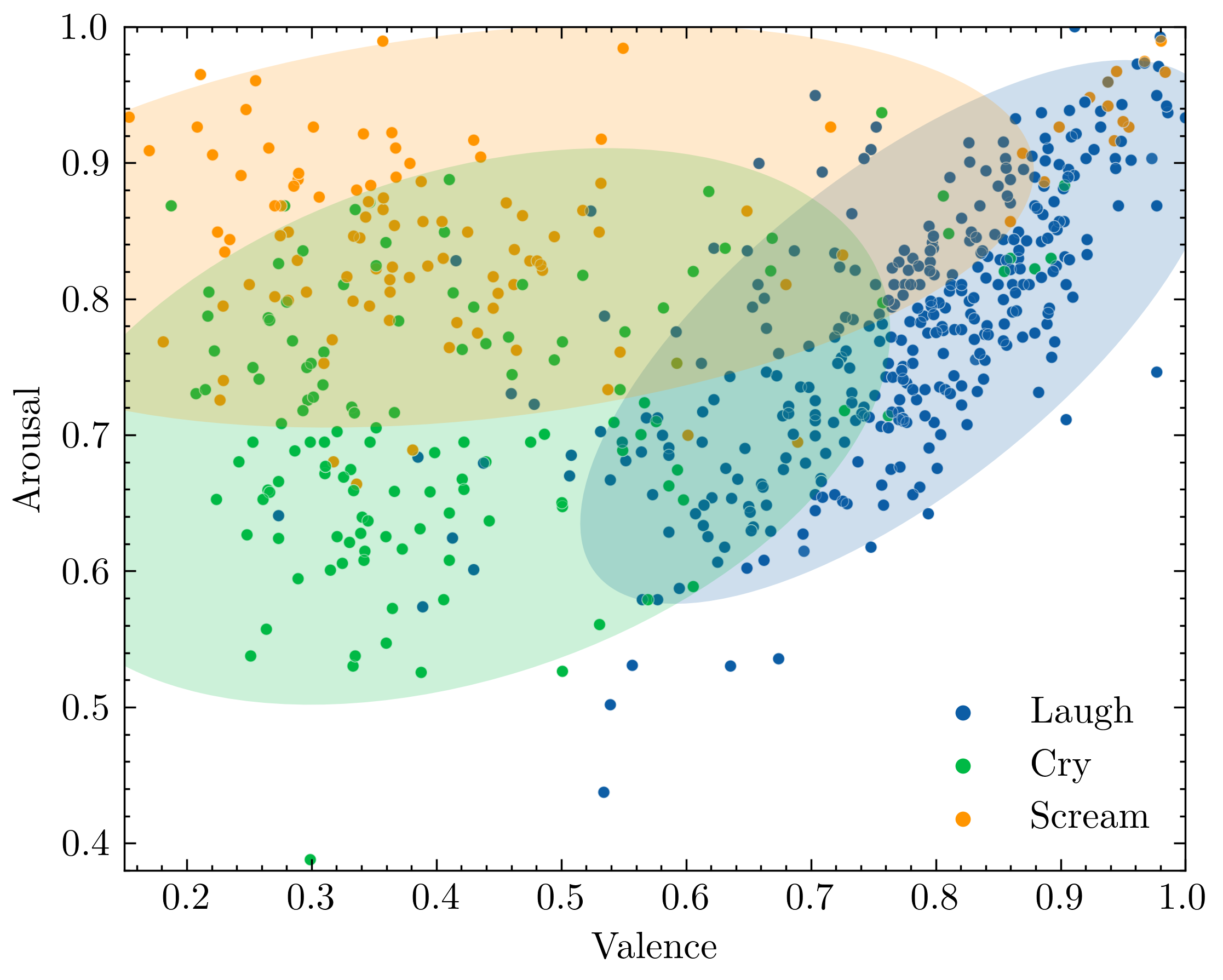}}%  \vspace{2.0cm}
\end{minipage}
\vspace{-1em}
\caption{Distribution of \textit{laugh}, \textit{cry} and \textit{scream} of the TYPE task in the arousal-valence space of the TWO task. It can be found that different VB types have different distributions that could be modeled.}
\label{fig:low_high}
\vspace{-1em}
\end{figure}

\subsection{Hierarchical Multi-task Learning}
% shaqra2019recognizing (hierarchical)
% russell1980circumplex (2->high)
% anuchitanukul2022burst2vec (shared specific)
%The whole hierarchical multi-task framework is illustrated in Fig.~\ref{fig:overview}. 
We propose an elaborate hierarchical framework to explicitly model the relationships between the tasks. There are five tasks investigated in our framework \cite{baird2022acii}:
\begin{itemize}
    \item \textbf{TWO} This task aims to predict the emotion of AB in a space with two dimensions, i.e., arousal and valence, based on the circumplex model of affect \cite{russell1980circumplex}.
    \item \textbf{HIGH} The HIGH task is to predict the emotion intensity in a higher-dimensional space of 10 emotion classes, including  \textit{surprise}, \textit{sadness}, \textit{excitement}, \textit{fear}, etc.
    \item \textbf{COUNTRY} We design this classification task to consider the relationship between VB and habitation locations. There are 4 countries considered in this task, i.e., U.S., China, Venezuela and South Africa.
    \item \textbf{CULTURE} This is a 10-dimensional, 4-country culture-specific emotion intensity regression task.
    \item \textbf{TYPE} This task focuses on the prediction of 8 VB types, i.e., \textit{cry}, \textit{gasp}, \textit{groan}, \textit{grunt}, \textit{laugh}, \textit{pant}, \textit{scream}, and \textit{other}.
\end{itemize}
Following \cite{anuchitanukul2022burst2vec}, we used different layers to disentangle task-agnostic and task-specific information. The shared feature extractor is trained to extract features that are generally useful for the different prediction tasks, while each task-specific feature extractor captures information that is more related to the corresponding task.

In terms of the relationship between emotion dimensions, 
%From the perspective of emotion dimension, the labels of the TWO, HIGH and CULTURE tasks are from low to high dimensions. Specifically, 
the arousal and valence values in the TWO task can imply the emotion classes in the high-dimensional emotion space in the HIGH or TYPE tasks \cite{schubert1999measuring}. As shown in Fig.~\ref{fig:low_high}, distributions of the VB types, \textit{laugh}, \textit{cry} and \textit{scream}, are different in the arousal-valence space.
The labels in the CULTURE task are combinations of emotions and countries. Therefore, the predictions of the HIGH, TYPE, CULTURE and COUNTRY tasks are conditioned on the predicted results of the TWO task, i.e., the predicted arousal and valence values. Since CULTURE task targets are combinations of emotion classes and countries, the system is designed to generate the CULTURE outputs based on the predictions from the HIGH and the COUNTRY tasks.
%Therefore, we use the prediction from the TWO task to all the other tasks, the prediction from Country and TYPE to the HIGH and CULTURE tasks, and the prediction from HIGH to the CULTURE task.

\subsection{Bi-directional Regression Chain}
\begin{figure}[htb]
\centering
\vspace{-2em}
\begin{minipage}[b]{0.9\linewidth}
    \centerline{\includegraphics[width=\linewidth]{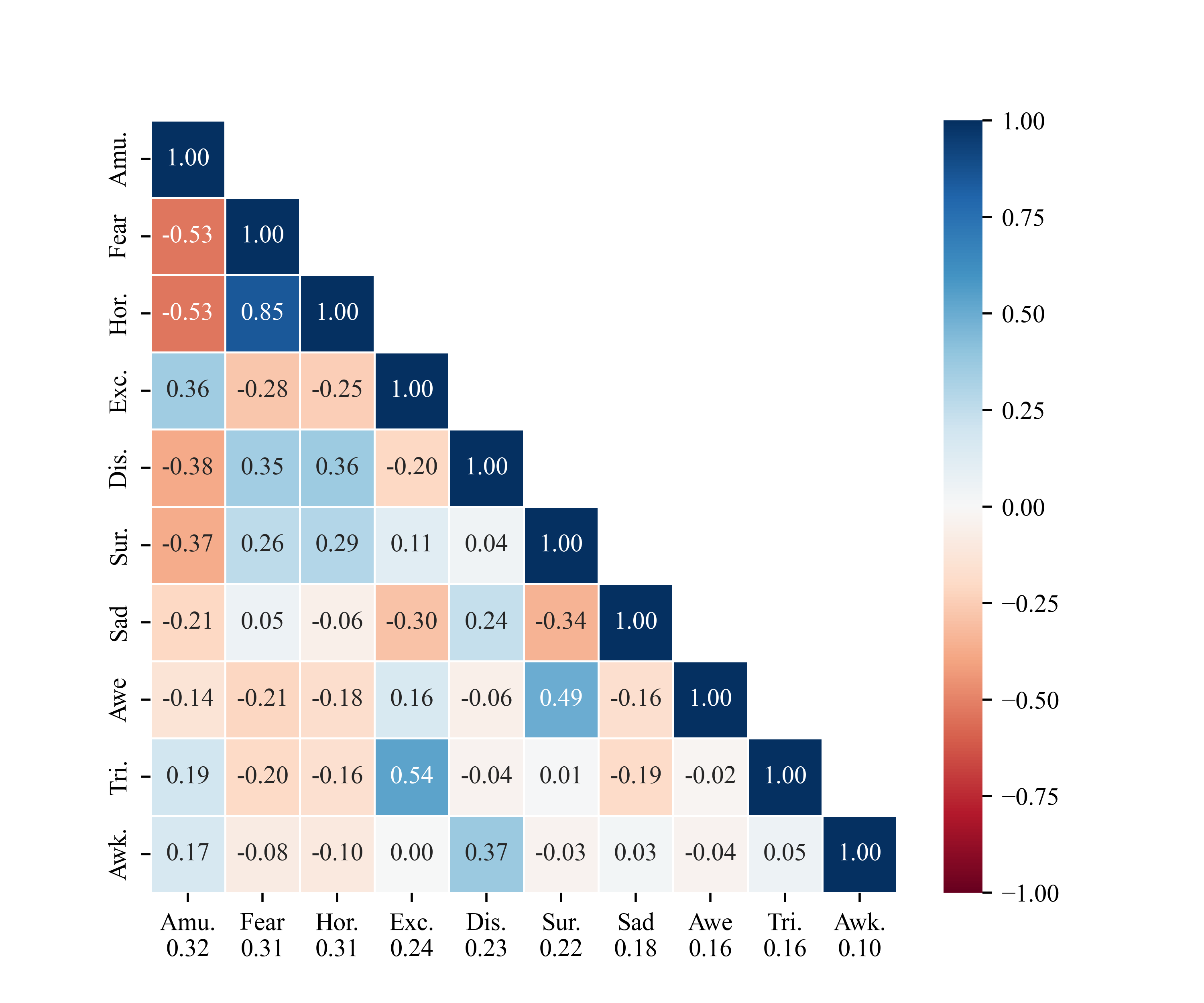}}%  \vspace{2.0cm}
\end{minipage}
\vspace{-1em}
\caption{Pearson correlation coefficients between emotion classes in the HIGH task based on training data.}
\label{fig:high_corr}
\end{figure}

It is noteworthy that the emotion classes are not independent, for example, a higher score in \textit{amusement} implies higher score in \textit{excitement} and lower scores in \textit{fear} and \textit{horror}.
We visualize the Pearson correlation coefficients between the emotion classes on the training subset in Fig.~\ref{fig:high_corr}. It is clearly shown that some pairs demonstrate significant correlation, which needs to be explicitly considered.

To model such relationships between emotion classes, we used a bi-directional regression chain to explicitly model the label dependency for the HIGH and CULTURE tasks. In a regression chain, with the predictor of the $i$-th emotion and the extracted feature denoted as $f_i$ and $z$, respectively, the emotion score is calculated by: $\hat{y}_i=\sigma(f_i(z\bigoplus \hat{y}_{<i}))$, where $\sigma$ is the sigmoid function, $\bigoplus$ is the vector concatenation operator, and $\hat{y}_{<i}$ is the previous predicted emotion scores before the \textit{i}-th emotion prediction.

To mitigate the effect from the emotion order of the chain, we accumulate absolute coefficients of each emotion, and arrange the order from higher accumulated values to lower ones, as shown in the x-axis in Fig.~\ref{fig:high_corr}. We then modify the chain regression layer to be bi-directional by adding another chain in the reverse direction.

% Following Xin \textit{et al.}~\cite{xin2022exploring}, we 

\subsection{Loss Functions}

For the countries and labels in TYPE task, categorical cross entropy (CE) is used as the main loss function. For the labels in the TWO, HIGH and CULTURE tasks, averaged biased concordance correlation coefficient (CCC) is adopted as the main loss function \cite{lawrence1989concordance}. The biased CCC is defined in Eq.~\ref{eq:ccc}.
\begin{equation}
\label{eq:ccc}
CCC(x_i, y_i) =\frac{1}{N}\sum\frac{2*cov(x_i, y_i)}{\sigma^{2}_{x_i} + \sigma^{2}_{y_i} + (\mu_{x_i} + \mu_{y_i})^2},
\end{equation}
where $N$ is the number of labels, and $\mu_{x_i}, \mu_{y_i}$ is the mean of $i$-th prediction and the corresponding ground truth values, respectively. The biased covariance is defined as $cov(x_i, y_i) = \sum(x_i - \mu_{x_i})(y_i - \mu_{y_i})$.

% \textcolor{red}{How do you combine the cross entorpy loss and the CCC loss??? weights??}
% Since multi-task learning can efficiently model the relationships between the labels, 
The total loss function is a weighted combination of the losses in the main task and the auxiliary tasks:
\begin{equation}
\label{eq:loss}
\mathcal{L}_{Target} = \lambda \mathcal{L}_{Target} + (1 - \lambda) * \sum(\mathcal{L}_{Auxiliary}),
\end{equation}
where $\mathcal{L}_{Target}$ and $\mathcal{L}_{Auxiliary}$ are the loss functions for the target and the auxiliary tasks, respectively, and $\lambda$ is a hyperparameter of the loss weights.

\section{Experiments}
\label{sec:exp}

\subsection{The A-VB Data}
We use the HUME-VB dataset of emotional non-linguistic vocalizations (vocal bursts)~\cite{Cowen2022HumeVB} that is used in the ACII A-VB Competition 2022 \cite{BairdA-VB2022}. The competition aims to promote research on modeling emotion in vocalizations, and proposes four tasks utilizing the HUME-VB data: the Two-Dimensional (TWO), High-Dimensional (HIGH), Cross-Cultural High-Dimensional (CULTURE) regression tasks, and the Expressive Burst-Type (TYPE) classification task.
The HUME-VB data contains about 37 hours of vocal burst data from 1702 speakers from China, South Africa, the U.S., and Venezuela. Each vocal burst is labeled with intensities in [1:100] of ten different expressed emotions or category in 8 classes from an average of 85.2 raters. The data is subsequently partitioned into training (19,990 VBs from 571 speakers), validation (19,396 VBs from 568 speakers), and test (19,815 VBs from 563 speakers) splits, with consideration of speaker independence and balances across countries and vocalization types.
% , as shown in Table ~\ref{tab:data}.
% \input{tables/data}

In this work, our target tasks are the TWO, HIGH and CULTURE tasks, while the classifications of COUNTRY and TYPE are used as auxiliary tasks. The TWO task aims to predict values of arousal and valence (based on 1=unpleasant/subdued, 5=neutral, 9=pleasant/stimulated), while The HIGH task aims to predict a higher dimension, i.e., the intensity of the aforementioned 10 emotions. The CULTURE task is a 10-dimensional, 4-country culture-specific emotion intensity regression task, i.e., it aims to predict the 40 intensity values of emotion (10 from each culture).

% \input{tables/data}
% \vspace{-1em}

\subsection{Experimental Setup}
% \textcolor{red}{Introduce the parameters for preprocessing (feature extraction window length, shift, augmentation parameters), parameters of attentive pooling (width), projection layer structure, shared layer, task-specific layers.
% Baseline systems introduction}
In this work, we set the dimensions of projection and shared layers to 128 and 64, respectively. The task-specific Bi-directional chains consist of two linear layers with sigmoid activation that are concatenated and averaged. The $\lambda$ in Eq.~\ref{eq:loss} is set to 0.9. We use  AdamW~\cite{loshchilov2017decoupled} as our optimizer with a learning rate of $1e-5$ for the Wav2vec 2.0 model finetuning and $1e-3$ for the downstream module training. To obtain a stabler CCC loss and alleviate the variance from the large pretrained model, we train the system with a large batch size of 1024 and a weight decay of $1e-3$. A 0.25 dropout is added between every two modules. We also apply early stopping (patience of 10, maximum of 25 epochs) to avoid overfitting the model.
The systems are evaluated on the validation and test datasets with the averaged biased CCC metric for the target tasks.

\subsection{Baselines}
The baseline systems in this challenge include feature-based and end-to-end methods~\cite{BairdA-VB2022}. The feature-based approach extracts 6,373-dimensional ComParE~\cite{schuller2013interspeech}, 88-dimensional eGeMAPS~\cite{eyben2015geneva} acoustic feature sets, and models the features with three fully-connected layers with layer normalization. While the end-to-end approach uses Emo-18~\cite{tzirakis2018end} convolutional neural networks followed by a 2-layer Long-short term memory (LSTM) network.

\subsection{Experimental Results}
% performance of each task
% Cross Comparison
\begin{table}[htb]
\centering
\resizebox{\linewidth}{!}{
\begin{tabular}{l | r r | r r | r r }
\toprule
\multirow{2}{*}{Approach}
 & \multicolumn{2}{c|}{TWO} & \multicolumn{2}{c|}{HIGH} & \multicolumn{2}{c}{CULTURE} \\
        & Val.   & Test      & Val.       & Test      & Val.   & Test      \\
\midrule
ComParE  & .4942  & .4986  & .5154  & .5214  & .3867  & .3887 \\
eGeMAPS  & .4114  & .4143  & .4484  & .4496  & .3229  & .3214 \\
END2YOU  & .4988  & .5084  & .5638  & .5686  & .4359  & .4401 \\
\midrule
Ours     & \bf{.6966}  & \bf{.6854} & \bf{.7351}  & \bf{.7237} & \bf{.6464}  & \bf{.6017} \\
\bottomrule
\end{tabular}
}
\caption{Experimental results on the TWO, HIGH, and CULTURE tasks of the ACII A-VB challenge 2022. the mean concordance correlation coefficient (CCC) is reported.}
\label{tab:results}
\end{table}

% Val.
% TWO: 0.69655, {'Valence': .7622, 'Arousal': .6309}
% HIGH: 0.73511, ['Awe', 'Excitement', 'Amusement', 'Awkwardness', 'Fear', 'Horror', 'Distress', 'Triumph', 'Sadness', 'Surprise'] [0.8169, 0.6962, 0.7928, 0.6085, 0.7742, 0.7528, 0.701, 0.6914, 0.711, 0.8063]
% CULTURE: 0.64641, ['China_Awe', 'China_Excitement', 'China_Amusement', 'China_Awkwardness', 'China_Fear', 'China_Horror', 'China_Distress', 'China_Triumph', 'China_Sadness', 'China_Surprise', 'United States_Awe', 'United States_Excitement', 'United States_Amusement', 'United States_Awkwardness', 'United States_Fear', 'United States_Horror', 'United States_Distress', 'United States_Triumph', 'United States_Sadness', 'United States_Surprise', 'South Africa_Awe', 'South Africa_Excitement', 'South Africa_Amusement', 'South Africa_Awkwardness', 'South Africa_Fear', 'South Africa_Horror', 'South Africa_Distress', 'South Africa_Triumph', 'South Africa_Sadness', 'South Africa_Surprise', 'Venezuela_Awe', 'Venezuela_Excitement', 'Venezuela_Amusement', 'Venezuela_Awkwardness', 'Venezuela_Fear', 'Venezuela_Horror', 'Venezuela_Distress', 'Venezuela_Triumph', 'Venezuela_Sadness', 'Venezuela_Surprise'] [0.3774, 0.6387, 0.5171, 0.5172, 0.643, 0.759, 0.6761, 0.6146, 0.7021, 0.7042, 0.784, 0.7051, 0.8387, 0.6326, 0.7269, 0.7092, 0.6927, 0.7134, 0.7396, 0.76, 0.626, 0.6917, 0.7429, 0.5253, 0.7041, 0.6472, 0.583, 0.6342, 0.6499, 0.7157, 0.7036, 0.4383, 0.7346, 0.5048, 0.5992, 0.6287, 0.45, 0.5477, 0.6815, 0.5964]

We compare our system with the baselines on the TWO, HIGH and CULTURE tasks in Table~\ref{tab:results}. It can be found that the proposed system outperforms the baselines on all three tasks by a significant margin. This demonstrates the effectiveness of the proposed hierarchical framework. 

\begin{table}[htb]
\centering
% \resizebox{\linewidth}{!}{
\begin{tabular}{l | c }
\toprule
Approach & Averaged CCC\\
\midrule
ComParE & .5154 \\
eGeMAPS & .4484 \\
END2YOU & .5638 \\
\midrule
Ours     & .7351\\
- Finetune & .6103\\
- Regression Chain & .6513\\
- Finetune \& Regression Chain & .5540\\
\bottomrule
\end{tabular}
% }
\caption{Ablation study for the HIGH task on the validation data. ``Ours" means the hierarchical framework with chain regression and finetuned Wav2vec 2.0, ``-" means removing corresponding module from ``Ours".}
\label{tab:ablation}
\end{table}

We also conducted experiments to verify the effectiveness of the integrated pre-trained representations and the regression chains on the HIGH task. As shown in Table~\ref{tab:ablation}, when the SSL representations are directly used without further fine-tuning on the HUME-VB dataset, the performance drops from 0.7351 to 0.6103, but still outperforms the baseline systems. If the regression chains are removed, the performance also decreases significantly, which demonstrates the effectiveness of the regression chains for the HIGH task. These results also suggest that the combination of fine-tuned SSL representations that implicitly borrow from external data, and the regression chains that model interactions between emotion classes, are both beneficial for performance.  

\begin{table}[htb]
\centering
% \resizebox{\linewidth}{!}{
\begin{tabular}{l | c c | c }
\toprule
Approach & Valence & Arousal & Average \\
\midrule
Ours     & .7622  & .6309 & .6966 \\
\bottomrule
\end{tabular}
% }
\caption{Performance (CCC) of proposed system for the TWO task on all validation data.}
\label{tab:two}
\end{table}

We further analyze the performance of the arousal and valence prediction in the TWO task. The breakdown of performance is shown in Table~\ref{tab:two}. It can be observed that the CCC of predicted valence values is much higher than that of predicted arousal values. This matches well with the characteristics of the HUME-VB dataset -- that the distribution of human valence annotation is more diffuse than the arousal distribution \cite{baird2022acii}. 

\begin{table}[htb]
\centering
\resizebox{\linewidth}{!}{
\begin{tabular}{l | c c c c c }
\toprule
Approach & Awe & Excite. & Amuse. & Awkward. & Fear \\
Ours & .8169 & .6962 & .7928 & .6085 & .7742 \\
\midrule
Approach & Horror & Distress & Triumph & Sadness & Surprise \\
Ours & .7528 & .7010 & .6914 & .7110 & .8063 \\
\bottomrule
\end{tabular}
}
\caption{Performance (CCC) of the proposed method for the HIGH task on the validation data.}
\label{tab:high}
\end{table}

For the HIGH task, the performances of different emotion classes are shown in Table~\ref{tab:high}. It can be seen that all 10 classes have satisfactory performance.  In particular, the \textit{awkward} class is relatively more difficult with a slightly lower performance, which is also observed in \cite{xin2022exploring}.

\begin{table}[htb]
\centering
% \resizebox{\linewidth}{!}{
\begin{tabular}{l | c | c c  }
\toprule
Countries  & Average & Train  & Val.  \\
\midrule
China    & .6149  & 79 & 76  \\
U.S.     & .7302 & 206 & 206 \\
South Africa & .6520 & 244 & 244 \\
Venezuela & .5885 & 42 & 42 \\
\bottomrule
\end{tabular}
% }
\caption{Performance (CCC) of the proposed method for the CULTURE task on the validation dataset. Distribution of recording numbers for the four countries on the training and validation sets is also shown.}
\label{tab:country}
\end{table}
In the CULTURE task, it can be found that the performance for the data from Venezuela is significantly worse than the other locations. This is probably caused by the unbalanced distribution in the dataset.  This is shown in Table~\ref{tab:country}, where the training and validation data for Venezuela is much less compared to the data for U.S. and South Africa. Similarly, the performance for China is also inferior to the those for U.S. and South Africa.
%It can be observed that the classes of \textit{excitement} and \textit{distress} have relatively lower performance.
% points (obs from table)

% \textcolor{red}{labels dependency (figs: type_two, high_corr)}
% insights (label dependency)
% figs/type_two: the relationship between low-dimensional labels (valence, arousal) and part of high-dimensional voc_type labels
% figs/high_corr: Pearson's correlation coefficients among 10 high-dimensional labels

% \textcolor{red}{Ablation Study (-Chain, -Hierarchical)}

\section{Conclusion}
\label{sec:conclusion}

In this paper, we investigate affective vocal burst recognition (AVBR) by proposing a hierarchical framework with bi-directional regression chains to explicitly consider multiple relationships, (i) between emotional states and diverse cultures, (ii) between low-dimensional and high-dimensional emotion spaces, and (ii) between various emotion classes within the high-dimensional space. To address the data sparsity problem in AVBR, we also integrate SSL representations via a trainable aggregation method. The proposed framework achieves significantly better performance than baseline systems on the HUME-VB dataset. Data analysis on the dataset and the experimental results also supports the necessity of modeling the inherent relationships.
In the future, we will investigate imbalanced learning w.r.t. cultures and labels in the AVBR task. We will also try to interpret the affective cues from the high-level embeddings for VBs.
% In the future, we will conduct research in low-resource VB recognition in cultures with limited training data.
% \section{Ethical Impact Statement}

% The proposed models may positively impact psychological assessments, human-computer interactions, etc. But they may be biased in gender or culture due to data imbalance.

% \section{Acknowledgment}

% This work was supported by Microsoft Azure Machine Learning Studio.

% 双向 v.s. 单向 (chain)
% Attn v.s. Mean
% Finetune
% 相关性(high -> two, high)

% references
% \balance
\bibliographystyle{IEEEbib}
\bibliography{main}

\begin{thebibliography}{10}

\bibitem{cowen2019mapping}
A.~S. Cowen, H.~A. Elfenbein, et~al.,
\newblock ``Mapping 24 emotions conveyed by brief human vocalization.,''
\newblock {\em American Psychologist}, 2019.

\bibitem{hawk2009worth}
S.~T. Hawk, G.~A. Van~Kleef, et~al.,
\newblock ``"worth a thousand words``: absolute and relative decoding of
  nonlinguistic affect vocalizations.,''
\newblock {\em Emotion}, 2009.

\bibitem{liang2021multibench}
P.~P. Liang, Y. Lyu, et~al.,
\newblock ``Multibench: Multiscale benchmarks for multimodal representation
  learning,''
\newblock {\em arXiv preprint arXiv:2107.07502}, 2021.

\bibitem{li2022context}
J. Li, S. Wang, et~al.,
\newblock ``Context-aware multimodal fusion for emotion recognition,''
\newblock {\em INTERSPEECH}, 2022.

\bibitem{singh2022systematic}
Y.~B. Singh and S. Goel,
\newblock ``A systematic literature review of speech emotion recognition
  approaches,''
\newblock {\em Neurocomputing}, 2022.

\bibitem{huang2019speech}
K.-Y. Huang, C.-H. Wu, et~al.,
\newblock ``Speech emotion recognition using deep neural network considering
  verbal and nonverbal speech sounds,''
\newblock in {\em ICASSP}. IEEE, 2019.

\bibitem{hsu2021speech}
J.-H. Hsu, M.-H. Su, et~al.,
\newblock ``Speech emotion recognition considering nonverbal vocalization in
  affective conversations,''
\newblock {\em IEEE/ACM TASLP}, 2021.

\bibitem{baird2022icml}
A. Baird, P. Tzirakis, et~al.,
\newblock ``The icml 2022 expressive vocalizations workshop and competition:
  Recognizing, generating, and personalizing vocal bursts,''
\newblock {\em arXiv preprint arXiv:2205.01780}, 2022.

\bibitem{baird2022acii}
A. Baird, P. Tzirakis, et~al.,
\newblock ``The acii 2022 affective vocal bursts workshop \& competition:
  Understanding a critically understudied modality of emotional expression,''
\newblock {\em arXiv preprint arXiv:2207.03572}, 2022.

\bibitem{jing2022redundancy}
X. Jing, M. Song, et~al.,
\newblock ``Redundancy reduction twins network: A training framework for
  multi-output emotion regression,''
\newblock {\em arXiv preprint arXiv:2206.09142}, 2022.

\bibitem{song2022dynamic}
M. Song, Z. Yang, et~al.,
\newblock ``Dynamic restrained uncertainty weighting loss for multitask
  learning of vocal expression,''
\newblock {\em arXiv preprint arXiv:2206.11049}, 2022.

\bibitem{baevski2020wav2vec}
A. Baevski, Y. Zhou, et~al.,
\newblock ``wav2vec 2.0: A framework for self-supervised learning of speech
  representations,''
\newblock {\em Advances in Neural Information Processing Systems}, 2020.

\bibitem{sharma2022multi}
M. Sharma,
\newblock ``Multi-lingual multi-task speech emotion recognition using wav2vec
  2.0,''
\newblock in {\em ICASSP}. IEEE, 2022.

\bibitem{purohit2022comparing}
T. Purohit, I.~B. Mahmoud, et~al.,
\newblock ``Comparing supervised and self-supervised embedding for exvo
  multi-task learning track,''
\newblock {\em arXiv preprint arXiv:2206.11968}, 2022.

\bibitem{pepino2021emotion}
L. Pepino, P. Riera, and L. Ferrer,
\newblock ``Emotion recognition from speech using wav2vec 2.0 embeddings,''
\newblock {\em INTERSPEECH}, 2021.

\bibitem{liu2020temporal}
J. Liu, Z. Liu, et~al.,
\newblock ``Temporal attention convolutional network for speech emotion
  recognition with latent representation.,''
\newblock in {\em INTERSPEECH}, 2020.

\bibitem{wen2021crossmodal}
H. Wen, S. You, and Y. Fu,
\newblock ``Cross-modal dynamic convolution for multi-modal emotion
  recognition,''
\newblock {\em Journal of Visual Communication and Image Representatio}, 2021.

\bibitem{chochlakis2022leveraging}
G. Chochlakis, G. Mahajan, et~al.,
\newblock ``Leveraging label correlations in a multi-label setting: A case
  study in emotion,''
\newblock {\em arXiv preprint arXiv:2210.15842}, 2022.

\bibitem{cariani1996neural}
P.~A. Cariani and B. Delgutte,
\newblock ``Neural correlates of the pitch of complex tones. ii. pitch shift,
  pitch ambiguity, phase invariance, pitch circularity, rate pitch, and the
  dominance region for pitch,''
\newblock {\em Journal of neurophysiology}, 1996.

\bibitem{colosi1998efficient}
J.~A. Colosi and M.~G. Brown,
\newblock ``Efficient numerical simulation of stochastic internal-wave-induced
  sound-speed perturbation fields,''
\newblock {\em The Journal of the Acoustical Society of America}, 1998.

\bibitem{adoma2020comparative}
A.~F. Adoma, N.-M. Henry, and W. Chen,
\newblock ``Comparative analyses of bert, roberta, distilbert, and xlnet for
  text-based emotion recognition,''
\newblock in {\em Proc. ICCWAMTIP}. IEEE, 2020.

\bibitem{wang2021fine}
Y. Wang, A. Boumadane, and A. Heba,
\newblock ``A fine-tuned wav2vec 2.0/hubert benchmark for speech emotion
  recognition, speaker verification and spoken language understanding,''
\newblock {\em arXiv preprint arXiv:2111.02735}, 2021.

\bibitem{conneau2020unsupervised}
A. Conneau, A. Baevski, et~al.,
\newblock ``Unsupervised cross-lingual representation learning for speech
  recognition,''
\newblock {\em arXiv preprint arXiv:2006.13979}, 2020.

\bibitem{ardila2019common}
R. Ardila, M. Branson, et~al.,
\newblock ``Common voice: A massively-multilingual speech corpus,''
\newblock {\em arXiv preprint arXiv:1912.06670}, 2019.

\bibitem{santos2016attentive}
C.~d. Santos, M. Tan, et~al.,
\newblock ``Attentive pooling networks,''
\newblock {\em arXiv preprint arXiv:1602.03609}, 2016.

\bibitem{ioffe2015batch}
S. Ioffe and C. Szegedy,
\newblock ``Batch normalization: Accelerating deep network training by reducing
  internal covariate shift,''
\newblock in {\em International conference on machine learning}. PMLR, 2015.

\bibitem{russell1980circumplex}
J.~A. Russell,
\newblock ``A circumplex model of affect.,''
\newblock {\em Journal of personality and social psychology}, 1980.

\bibitem{anuchitanukul2022burst2vec}
A. Anuchitanukul and L. Specia,
\newblock ``Burst2vec: An adversarial multi-task approach for predicting
  emotion, age, and origin from vocal bursts,''
\newblock {\em arXiv preprint arXiv:2206.12469}, 2022.

\bibitem{schubert1999measuring}
E. Schubert,
\newblock ``Measuring emotion continuously: Validity and reliability of the
  two-dimensional emotion-space,''
\newblock {\em Australian Journal of Psychology}, 1999.

\bibitem{lawrence1989concordance}
I. Lawrence and K. Lin,
\newblock ``A concordance correlation coefficient to evaluate
  reproducibility,''
\newblock {\em Biometrics}, 1989.

\bibitem{Cowen2022HumeVB}
A. Cowen, A. Bard, et~al.,
\newblock ``The hume vocal burst competition dataset {(H-VB)},''
\newblock {\em Zenodo}, 2022.

\bibitem{BairdA-VB2022}
A. Baird, P. Tzirakis, et~al.,
\newblock ``The {ACII} 2022 affective vocal bursts workshop and competition:
  Understanding a critically understudied modality of emotional expression,''
  2022.

\bibitem{loshchilov2017decoupled}
I. Loshchilov and F. Hutter,
\newblock ``Decoupled weight decay regularization,''
\newblock {\em arXiv preprint arXiv:1711.05101}, 2017.

\bibitem{schuller2013interspeech}
B. Schuller, S. Steidl, et~al.,
\newblock ``The interspeech 2013 computational paralinguistics challenge:
  Social signals, conflict, emotion, autism,''
\newblock in {\em INTERSPEECH}, 2013.

\bibitem{eyben2015geneva}
F. Eyben, K.~R. Scherer, et~al.,
\newblock ``The geneva minimalistic acoustic parameter set (gemaps) for voice
  research and affective computing,''
\newblock {\em IEEE transactions on affective computing}, 2015.

\bibitem{tzirakis2018end}
P. Tzirakis, J. Zhang, and B.~W. Schuller,
\newblock ``End-to-end speech emotion recognition using deep neural networks,''
\newblock in {\em ICASSP}. IEEE, 2018.

\bibitem{xin2022exploring}
D. Xin, S. Takamichi, and H. Saruwatari,
\newblock ``Exploring the effectiveness of self-supervised learning and
  classifier chains in emotion recognition of nonverbal vocalizations,''
\newblock {\em arXiv preprint arXiv:2206.10695}, 2022.

\end{thebibliography}

\end{document}